\documentclass[iop,numberedappendix,appendixfloats]{emulateapj}
\shortauthors{Yan \& Ma}

\usepackage{natbib}
\usepackage{enumitem}
\usepackage{amsmath}
\usepackage{soul}

\usepackage[bookmarks=true, pagebackref=true]{hyperref}
\usepackage[all]{hypcap}

\usepackage[load-configurations=astronomy]{siunitx}
\DeclareSIUnit\micron{\ensuremath{\mathrm{\micro\meter}}}
\DeclareSIUnit\sig{\ensuremath{\mathrm{\sigma}}}
\DeclareSIUnit\mag{mag}
\DeclareSIUnit\jansky{Jy}
\DeclareSIUnit\deg{deg}
\DeclareSIUnit\yr{yr}
\DeclareSIUnit\dex{dex}
\DeclareSIUnit\Lsun{\ensuremath{\mathrm{L_\sun}}}
\DeclareSIUnit\Msun{\ensuremath{\mathrm{M_\sun}}}

\newcommand{\Herschel}{\textit{Herschel}}

\begin{document}

\title{On the Limited Sizes of Dusty Starbursting Regions at High Redshifts}

\author{Haojing Yan and Zhiyuan Ma}
\affil{Department of Physics and Astronomy, University of Missouri-Columbia, USA}
\email{yanha@missouri.edu, zmzff@mail.missouri.edu}

\begin{abstract}

   Using the far-infrared data obtained by the {\it Herschel Space
Observatory}, we study the relation between the infrared luminosity ($L_{IR}$)
and the dust temperature ($T$) of dusty starbursting galaxies at high redshifts
(high-z). We focus on the total infrared luminosity from the cold-dust
component ($L_{IR}^{(cd)}$), whose emission can be described by a modified
black body (MBB) of a single temperature ($T_{mbb}$). An object on the
($L_{IR}^{(cd)}$, $T_{mbb}$) plane can be explained by the equivalent of the
Stefan-Boltzmann law for a MBB with an effective radius of $R_{eff}$. We show
that $R_{eff}$ is a good measure of the combined size of the dusty starbursting
regions (DSBRs) of the host galaxy. In at least one case where the individual
DSBRs are well resolved through strong gravitational lensing, $R_{eff}$ is
consistent with the direct size measurement. We show that the observed
$L_{IR}\text{--}T$ relation is simply due to the limited
$R_{eff}$ ($\lesssim 2$~kpc). The small $R_{eff}$ values also agree with the
compact sizes of the DSBRs seen in the local universe. However, previous
interferometric observations to resolve high-z dusty starbursting galaxies
often quote much larger sizes. This inconsistency can be reconciled by the
blending effect when considering that the current interferometry might still
not be of sufficient resolution. From $R_{eff}$ we infer the lower limits to
the volume densities of the star formation rate (``minSFR3D'') in the DSBRs,
and find that the $L_{IR}$-$T$ relation outlines a boundary on the
\big($L_{IR}^{(cd)}$, $T$\big) plane, below which is the ``zone of avoidance''
in terms of minSFR3D.

\end{abstract}


\section{Introduction}

   Dusty infrared (IR) galaxies are known to have a correlation between their
infrared luminosities and dust temperatures (``$L_{IR}\text{--}T$ relation'').
Generally speaking, ultra-luminous infrared galaxies (ULIRGs;
$L_{IR} \sim 10^{12-13}L_\odot$) and hyper-luminous infrared galaxies (HyLIRGs;
$L_{IR} \gtrsim 10^{13}L_\odot$) have typical dust temperatures of
$\sim 40\text{--}60$~K, while luminous infrared galaxies (LIRGs;
$L_{IR} \sim 10^{11-12}L_\odot$) and others at lower luminosities have lower
dust temperatures of $\sim 20\text{--}30$~K. This trend was clearly revealed
when sub-millimeter galaxies (SMGs), which are high-z ULIRGs, were compared
to ULIRGs and other IR galaxies in the local universe \citep[e.g.,][]{Blain2003,
Chapman2003}. There have been a number of studies to understand the nature of
this relation and its dispersions, such as its dependence on redshifts,
different galaxy populations, etc. \citep[e.g.][]{Chapman2005, Kovacs2006,
Clements2010, Magdis2010, Hwang2010, Symeonidis2013}.

  In \citet[hereafter MY15]{MY15}, we have studied the quasars from the Sloan
Digital Sky Survey (SDSS) that have far-IR (FIR) counterparts in the wide-field
survey data from the {\it Herschel}{} Space Observatory \citep{Pilbratt2010}.
We have shown that the majority of them are ULIRGs, and that their FIR emission
originated from the cold-dust component is predominantly powered by starbursts
rather than active galactic nuclei (AGN). One of our conclusions is that they
follow the same $L_{IR}\text{--}T$ relation. We have further shown that this
relation is simply due to the limited maximum size of the combined dusty
starbursting regions (hereafter DSBRs) in the host galaxies. 

   The motivation of our current work is that this maximum size is only around
$\sim 2$~kpc, which seems to be significantly smaller than the claimed physical
sizes of high-z ULIRGs based on a number of direct measurements through
sub-millimeter (sub-mm) or radio interferometry. These observations have
sub-arcsec resolutions, and typically result in rather extended sizes of
$\sim 4\text{--}8$~kpc or even larger
\citep[e.g.,][]{Chapman2004, Biggs2008, Younger2008, Younger2010, Simpson2015}.

   Here we further investigate this problem, using an enlarged sample that
incorporates different high-z IR galaxy populations. Throughout this
{\it Letter}, we adopt the following cosmological parameters:
$\Omega_M=0.27$, $\Omega_\Lambda=0.73$ and $H_0=71$~km/s/Mpc.

\section{Data and Analysis}

   We followed MY15 in this current analysis. Only the objects that have
spectroscopic redshifts were used. The FIR counterpart identification was
mainly based on the latest source catalogs from three major wide field
surveys by \Herschel{}, namely, the Herschel Multi-tiered Extragalactic Survey
\citep[HerMES, $\sim$\SI{100}{\deg\squared};][]{Oliver2012, Wang2013}, the
Herschel Stripe 82 Survey
\citep[HerS, $\sim$\SI{80}{\deg\squared};][]{Viero2014}, and the Scientific
Demonstration Phase data of the Herschel Astrophysical Terahertz Large Area
Survey \citep[H-ATLAS SDP, $\sim$\SI{19.3}{\deg\squared};][]{Eales2010, 
Ibar2010, Pascale2011, Rigby2011}.
The FIR spectral energy distributions (SEDs)
were constructed using the three-band photometry from the Spectral
and Photometric Imaging REceiver \citep[SPIRE;][]{Griffin2010} at 250, 350 and
\SI{500}{\micron}. We only considered the objects that are detected in all
these three bands.

\subsection{Input samples}

   The largest addition to the MY15 sample is through the use of the 
$13^{th}$ edition V{\'e}ron Catalogue of Qusars and Active Nuclei
\citep{Veron2010}, which contains \num{34231} and \num{133332} AGN and quasars,
respectively (hereafter ``V-AGN'' and ``V-QSO'', respectively). The next is 
from \citet[][``C12'']{Casey2012b}, which contains \num{767} spectroscopically
confirmed SPIRE sources in the HerMES fields. We also included the SMGs from
\citet[][``C05'']{Chapman2005} and \citet[][``I05'']{Ivison2005}, and a few
sources of mixed selections from \citet[][``M11'']{Magdis2011} and
\citet[][``Y14'']{Yan2014}. Any duplicates among these sources and/or the
MY15 sample were removed. As in MY15, we obtained the SPIRE photometry by
matching the sources to the aforementioned {\it Herschel}{} catalogs, adopting
the matching radius of \SI{3}{\arcsec} to minimize the contaminated sources
due to blending.

  We also incorporated a number of objects that are outside of the
aforementioned {\it Herschel}{} survey fields but have reported SPIRE 
photometry. These include the high-z quasars from
\citet[][``L13'']{Leipski2013},
the radio galaxies from \citet[][``D14'']{Drouart2014}, and the 3C
radio sources from \citet[][``P15'']{Podigachoski2015}. While these objects
certainly contain powerful AGN, we are convinced (as are these authors also
inclined to believe) that their FIR emissions should be dominated by
starbursts as in the SDSS quasars presented in MY15.

   Finally, we included a few high-z ULIRGs that are known to be amplified by
strong gravitational lensing, which all have spectroscopic redshifts, reported
SPIRE photometry and adopted amplification factors (``$\mu$''). These include
the most highly lensed SMG from \citet{Swinbank2010} and
\citet[][``S\&I10'']{Ivison2010},
the strong Planck source from \citet[][``F12'']{Fu2012}, the South Pole
Telescope's lensed galaxies from \citet[]{Weiss2013} and \citet[][``W\&H13'']
{Hezaveh2013},
and the lensed galaxies in the H-ATLAS SDP from \citet[][``B13''; only the two
sources with grade ``A'' lensing models and $\mu>10$ were used]{Bussmann2013}. 

\subsection{SED fitting and the final sample}

   As in MY15, we only studied the coldest dust component, which dominates the
FIR emission that is sampled by the SPIRE bands. While this will certainly
underestimate (by $\sim 0.1\text{--}0.2$~dex) the total IR luminosity that
consists of the contributions from other components of higher temperatures, the
simplification has the advantages that the associated luminosity can be safely
attributed to starbursts and that the dust temperature is uniquely
defined. Following MY15, we analyzed their FIR SEDs by fitting a
single-temperature, modified blackbody (MBB) spectrum. We briefly describe the
procedure below.

   We used the \texttt{cmcirsed} code of \citet{Casey2012} to perform the MBB
fitting. The MBB spectrum can be written as
\begin{equation}\label{eq:mbb}
\begin{split}
     S_{\lambda}(\lambda) & \equiv N \cdot I_{mbb}(\lambda)\\
                & = N\frac{1-\mathrm{e}^
         {-(\frac{\lambda_0}{\lambda})^{\beta}}}{1-\mathrm{e}^{-1}}
     \frac{(\frac{2hc^2}{\lambda^5})}{\mathrm{e}^{hc/(\lambda kT_{mbb})}-1} \,,
\end{split}
\end{equation}
where $T_{mbb}$ is the characteristic temperature of the MBB, $N$ is the
scaling factor that is related to the intrinsic luminosity, $\beta$ is the
emissivity (when $\beta=0$ Equation~\ref{eq:mbb} reduces to the form of a
black body), and $\lambda_0$ is the reference wavelength where the opacity is
unity. We adopted $\beta=1.5$ and $\lambda_0=\SI{100}{\micron}$. 
The total IR luminosity of the cold-dust
component, $L_{IR}^{(cd)}$, can be obtained as
\begin{equation}\label{eq:lir}
    L_{IR}^{(cd)}\equiv L_{IR}^{mbb}\equiv
    \int_{8\micron}^{1000\micron} S_\lambda(\lambda)\mathrm{d}\lambda
\end{equation}
by integrating the best-fit MBB model
$S_\lambda(\lambda)$ from 8 to \SI{1000}{\micron}. 

\begin{figure}[t]
\plotone{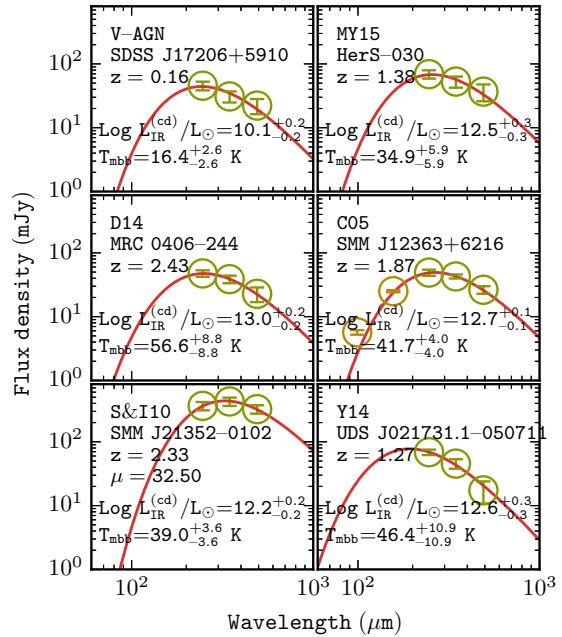}
\caption{Examples of FIR SED fitting, with the derived quantities labeled. The
data points are the photometry in SPIRE 250, 350 and \SI{500}{\micron},
respectively, while the curves are the best-fit MBB models. One of the objects
has photometry in two more bands bluer than the SPIRE bands, which were not
used in the fit but are plotted here to show that the fit based on only the
SPIRE bands indeed can get reliable results.
}
\label{fig:SEDfitting}
\end{figure}

\begin{figure*}[t]
\plotone{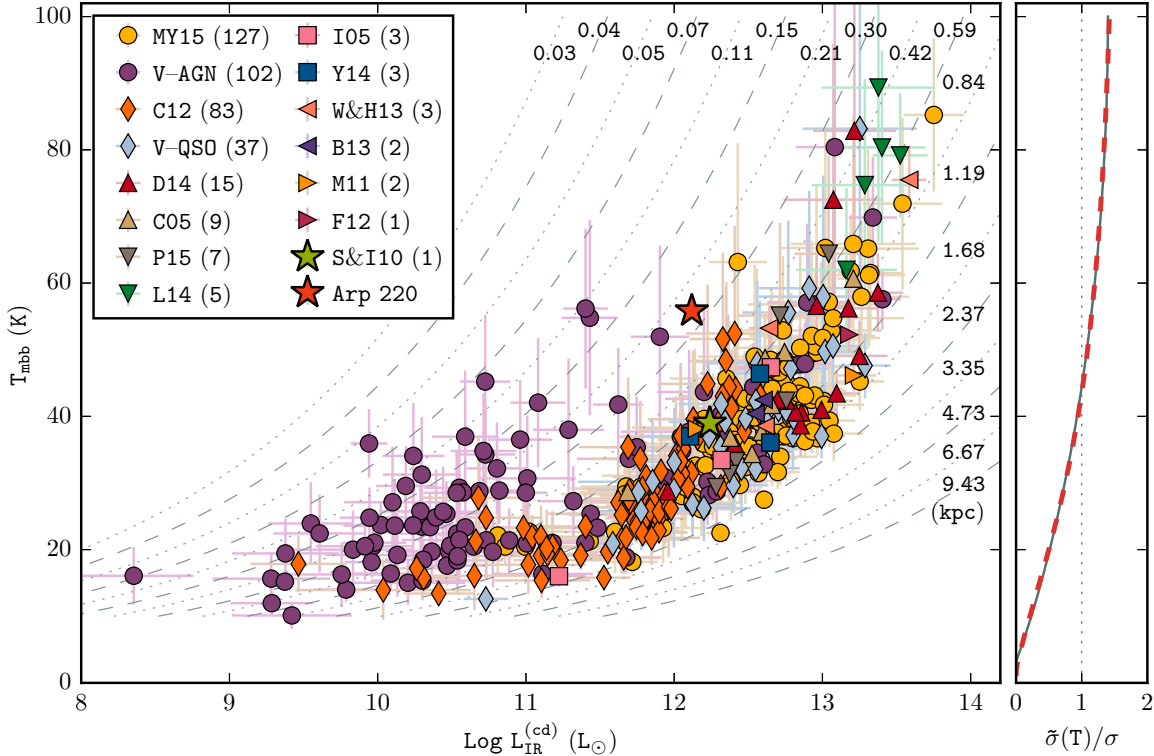}
\caption{$L_{IR}\text{--}T$ relation as revealed by our final sample. The
individual objects from different input samples and the total numbers are
shown in different symbols of colors as in the legend. The dashed curves
are a family of MBB S-B equivalent with different $R_{eff}$, whose values are
marked next to the curves (see \S 3.1). The right panel shows
$\tilde\sigma(T)/\sigma$ (red dashed curve) and the fit (grey solid curve).
}
\label{fig:LTobj}
\end{figure*}

   Our further analysis is based on $L_{IR}^{(cd)}$ and $T_{mbb}$ thus obtained.
We only retained the objects that have reasonable SED fitting quality
($\chi^2<10$) and reliable $T_{mbb}$ measurements
($T_{mbb}/\Delta T_{mbb}\geq 3$),
which sum up to \num{400} objects in total.
As examples, Figure~\ref{fig:SEDfitting} shows the SED fitting results for a
few objects from our final sample.
Figure~\ref{fig:LTobj} shows the $L_{IR}\text{--}T$ relation from the entire
sample, where the number of contributing objects from each initial sample is
also labeled. The distributions of redshifts and $L_{IR}^{(cd)}$ 
are shown in the first two histograms in Figure~\ref{fig:hist}.

\section{Interpretations}

   As addressed in MY15, the $L_{IR}\text{--}T$ trend revealed in
Figure~\ref{fig:LTobj} is not due to selection effects but is governed by the
equivalent of Stefan-Boltzmann law in case of MBB. Here we fully develop
this idea.

\subsection{Effective radius of DSBRs}

The location of an object on the \big($L_{IR}^{(cd)}$, $T_{mbb}$\big) plane is
determined by the intensity of the MBB as expressed in Equation~\eqref{eq:mbb}.
To obtain the total power radiated from the cold-dust component of the object,
i.e., its total luminosity, one should integrate the MBB function
$I_{mbb}(\lambda)$ over wavelength and over the solid angle (``$\Omega$'') that
its surface area (``A'') subtends as seen from the source, and multiply by 
this surface area:
\begin{equation}\label{eq:mbbsb1}
\begin{split}
    L & = A\cdot \int_0^\infty I_{mbb}(\lambda)\, \mathrm{d}\lambda
          \int \mathrm{d}\Omega\\
      & = A\cdot \frac{2\pi k^4}{h^3 c^2} T^4 \int_0^\infty \frac{1-\mathrm{e}^{-(\tau u)^\beta}}{1-\mathrm{e}^{-1}}
      \frac{u^3}{\mathrm{e}^u-1} \mathrm{d}u \,,
\end{split}
\end{equation}
where we make the substitution of $u=\frac{hc}{\lambda kT}$, set
$\tau\equiv \frac{\lambda_0 kT}{hc}$, and also write $T_{mbb}$ as $T$ for 
convenience. If there were no
the modified term $(1-e^{-(\tau u)^\beta})$ to the Planck's black body function,
the integral above would be $\pi^4/15$ and Equation~\eqref{eq:mbbsb1} would
reduce to the Stefan-Boltzmann law, i.e., $L=A\sigma T^4$, where 
$\sigma=\frac{2\pi^5K^4}{15h^3c^2}$ is the Stefan-Boltzmann constant. For
simplicity, we introduce
\begin{displaymath}
  \tilde{\sigma}(T,\beta,\lambda_0) = \frac{2\pi k^4}{h^3c^2}
      \int_0^\infty \frac{1-\mathrm{e}^{-(\tau u)^\beta}}{1-\mathrm{e}^{-1}} \frac{u^3}{\mathrm{e}^u-1} \mathrm{d}u \,,
\end{displaymath}
which is the equivalent of the Stefan-Boltzmann constant in the MBB case. Note
that $\tilde{\sigma}$ is dependent of $T$ because
$\tau\equiv \frac{\lambda_0 kT}{hc}$ is involved.

  In the context of this work, we can take the approximation that the integral
in Equation~\eqref{eq:mbbsb1} is only over the conventional total IR regime of 
$S_\lambda(\lambda)$ from 8 to \SI{1000}{\micron} as in Equation~\eqref{eq:lir}
such that $L\approx L_{IR}^{(cd)}$, i.e., we have
$L_{IR}^{(cd)}=A\tilde{\sigma}T_{mbb}^4$, which we shall refer to
as the ``MBB S-B equivalent''.

  Assuming spherical symmetry, we can define an ``effective radius'', $R_{eff}$,
such that $A=4\pi R_{eff}^2$.
This is to imagine that all the star-forming regions within the
galaxy are combined together and that the sum can be approximated by an
effective sphere with the radius of $R_{eff}$. We thus have
\begin{equation}\label{eq:mbbsb2}
   L_{IR}^{(cd)}=4\pi R_{eff}^2 \tilde{\sigma}T_{mbb}^4 \,.
\end{equation}
For $T_{mbb}$ within the range of interest ($\sim 10$--100~K), the deviation of
$\tilde{\sigma}$ from $\sigma$ is within a factor of two and can be well
approximated as
\begin{equation}\label{eq:SBsigratio}
  \tilde{\sigma}(T)/\sigma = 10^{-3}(-3.03T^{1.5}+45.55T-127.53) \,,
\end{equation}
for our choice of $\lambda_0=100$~$\mu$m and $\beta=1.5$ (see the right panel
of Figure~\ref{fig:LTobj}).

  Thus the data points on the \big($L_{IR}^{(cd)}$,$T_{mbb}$\big) plane
can be explained by a family of MBB S-B equivalent curves of different
$R_{eff}$, which are shown in Figure~\ref{fig:LTobj}.
A pair of \big($L_{IR}^{(cd)}$, $T_{mbb}$\big) values allow
us determine $R_{eff}$ of the combined dusty star forming regions in the
galaxy under question. MY15 uses the same argument, and shows that (1) in the
low luminosity regime the increasing of $L_{IR}^{(cd)}$ is due to the
increasing of $R_{eff}$, while in the high luminosity regime the increase of
$L_{IR}^{(cd)}$ has to be attributed to the increased heating intensity, and
(2) $R_{eff}$ cannot be increased arbitrarily and has a limit of 
$\lesssim 2$~kpc. MY15 takes a less rigorous approach and
approximates the temperature-dependent $\tilde{\sigma}$ by $\sigma T^\alpha$
and obtains $L_{IR}=4\pi R_{eff}^2 \sigma T^{4.32}$. The derivation of 
Equation~\eqref{eq:mbbsb2} here is more appropriate, which results in more
accurate determination of $R_{eff}$. 

\begin{figure*}[t]
\plotone{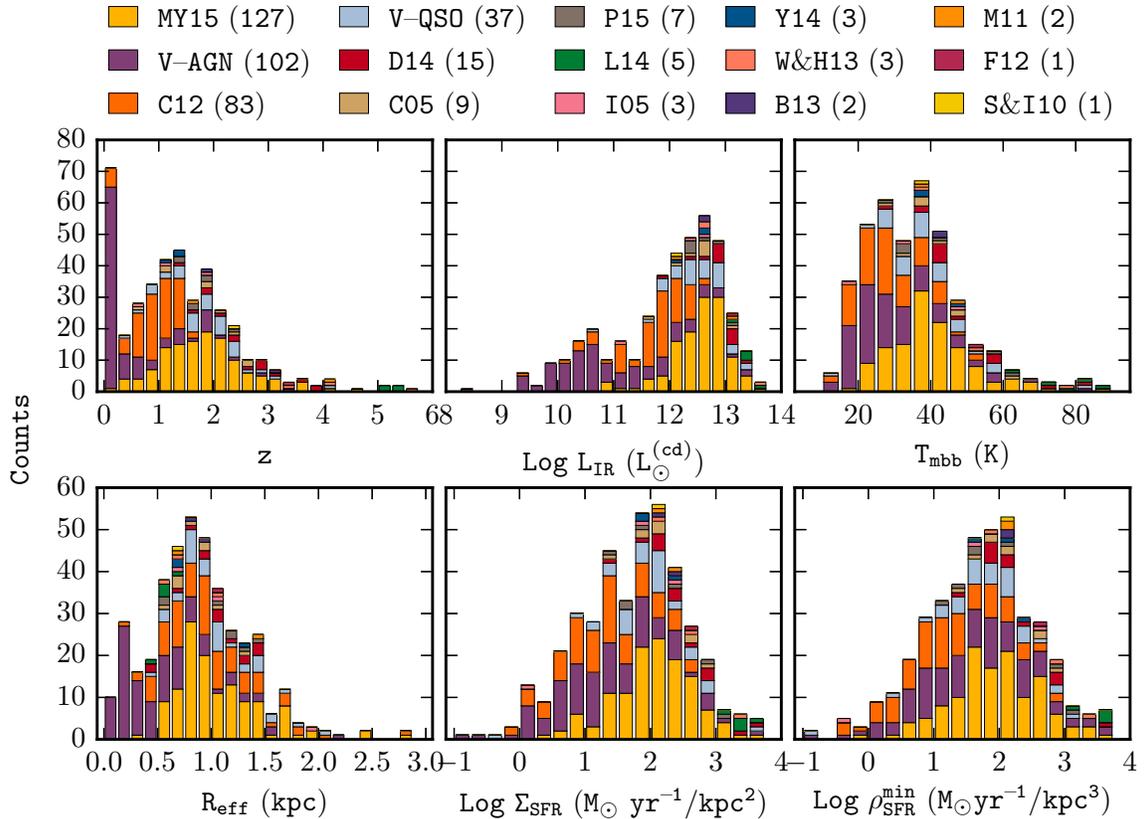}
\caption{Statistics of the objects from our sample.
}
\label{fig:hist}
\end{figure*}

   The histograms of $T_{mbb}$ and $R_{eff}$ are shown in Figure~\ref{fig:hist}
for all the objects in our sample. The vast majority of them (98.3\%)
have $R_{eff}\leq 2$~kpc, and this limit is the reason for the observed
$L_{IR}\text{--}T$ relation.

\subsection{SFR surface and volume densities}

   Following \citet{Kennicutt1998} to convert the infrared
luminosity to the SFR and assuming the initial mass function of
\citet{Chabrier2003}, we get
$\text{SFR}=\num{1.0d-10}L_{IR}^{(cd)}/L_\odot$~$\mathrm{M_\odot yr^{-1}}$.
From the $R_{eff}$ values above, we can derive the
SFR surface densities (hereafter ``SFR2D'', or ``$\Sigma_\text{SFR}$'') that are
often used in the literature, where the ``surface area'' going into the
calculation is the area of the DSBRs projected on the sky. Assuming spherical
symmetry, the total surface area of the $i$-th DSBR is $4\pi R_i^2$, where $R_i$
is its radius. We further make a highly simplified assumption that all the DSBRs
in a given galaxy have the same dust temperature. The definition of $R_{eff}$
thus implies that $4\pi R_{eff}^2=\sum\limits_{i} 4\pi R_i^2$, where the
summation over $i$ goes through all the DSBRs. The projected surface area of
the $i$-th DSBR on the sky is $\pi R_i^2$, and hence 
$\Sigma_\text{SFR}=\text{SFR}/\sum\limits_{i} \pi R_i^2=\text{SFR}/\pi R_{eff}^2$. Considering
Equation~\eqref{eq:mbbsb2} and that 
$\sigma=\num{1.411d5}L_\odot ~\mathrm{kpc^{-2} K^{-4}}$, we have
$\Sigma_\text{SFR}=\num{5.644d-5}(\tilde{\sigma}/\sigma)T_{mbb}^{4}$. This
means that $\Sigma_\text{SFR}$ is constant for a fixed $T_{mbb}$.

   Similarly, we can calculate the SFR volume density (hereafter ``SFR3D'', or
``$\rho_\text{SFR}$''). Under the same
assumptions as above, we define a different effective sphere with
the radius of $r_{eff}$, whose volume is equal to the sum of the volumes of the
individual DSBRs in a galaxy, i.e., 
$\frac{4}{3}\pi r_{eff}^3=\sum\limits_{i} \frac{4}{3}\pi R_i^3$. It is obvious
that $R_{eff} \geq r_{eff}$, and generally there is no easy way to infer
$r_{eff}$ from $R_{eff}$. However, in the limiting case that there is only one
DSBR in the galaxy under question, we should have $R_{eff}=r_{eff}$. Therefore,
we can calculate the {\it minimum} SFR volume density (hereafter ``minSFR3D'',
or ``$\rho_{\text{SFR}}^{min}$'') as
$\rho_{\text{SFR}}^{min}=\text{SFR}/(\frac{4}{3}\pi R_{eff}^3)$.

   The histograms of SFR2D and minSFR3D are shown in Figure~\ref{fig:hist}.

\subsection {Nature of the $L_{IR}$-$T$ relation}

\begin{figure*}[t]
\plotone{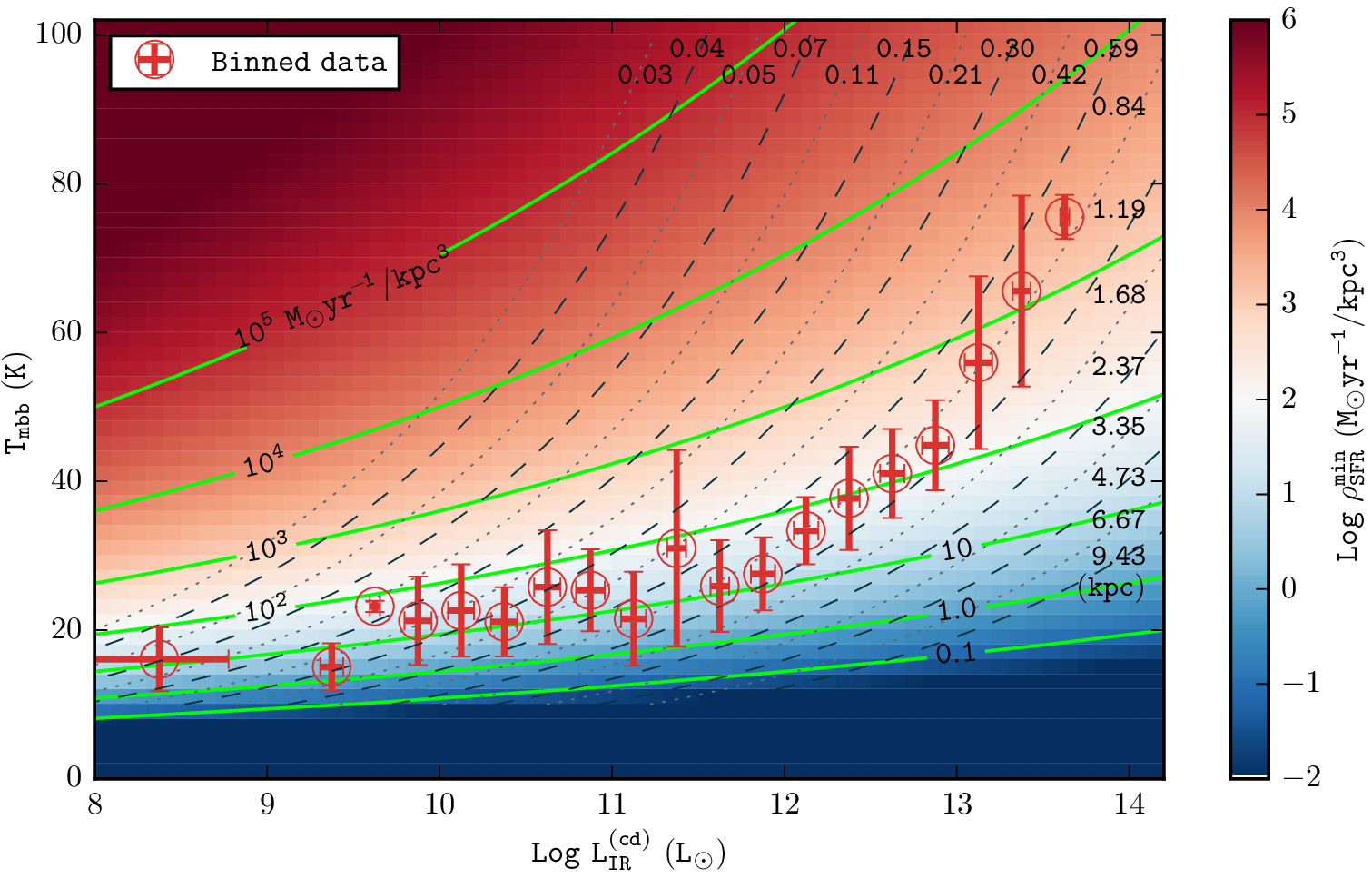}
\caption{Interpretation of the $L_{IR}\text{--}T$ relation in terms of minSFR3D.
The points are the averaged relation based on the data shown in
Figure~\ref{fig:LTobj}, binned in 0.2~dex in $L_{IR}^{cd}$. The grey dashed
curves are the same as in Figure~\ref{fig:LTobj}. The minSFRD3D
values on this plane are shown as the color-coded surface, and the green curves
mark a few representative values. 
}
\label{fig:TLIR_SFR3D}
\end{figure*}

   The $L_{IR}$-$T$ relation now has a new meaning. As
a pair of \big($L_{IR}^{(cd)}$, $T_{mbb}$\big) correspond to one $R_{eff}$
value, and hence one minSFR3D value, the $L_{IR}^{(cd)}$-$T$ plane can be
converted into a minSFR3D ``surface'', which is shown in
Figure~\ref{fig:TLIR_SFR3D}, where the average trend based on the data points
in Figure~\ref{fig:LTobj} is also displayed. It is immediately clear that our
objects form the observed $L_{IR}$-$T$ relation because they outline a region
that has a narrow spread in minSFR3D.

  We emphasize again that, as shown in the simulation of MY15 (see also \S 2
above), the lack of objects in the area below this relation cannot be due to
selection bias. From Figure~\ref{fig:TLIR_SFR3D}, it is clear that this
high-$L_{IR}^{(cd)}$, low-$T_{mbb}$ area corresponds to low minSFR3D. 
We therefore suggest that this area is a ``zone of avoidance'' in terms of
minSFR3D. In other words, very cold ULIRGs or HyLIRGs should be very rare, 
because they would require very large $R_{eff}$ (and hence very low minSFRD) in
order to achieve a high IR luminosity at a low dust temperature. On the other
hand, the lack of objects in the area above the current $L_{IR}$-$T$ relation
outlined by our objects {\it could be} due to the possible selection bias in our
current sample, where we limit to $S_{250}\geq 50$~mJy in order to include
the most reliable detections in the {\it Herschel}\, SPIRE bands. We refer the
readers to the discussion in MY15 for details (see \S 4.1 and Figure 12
therein).

\subsection {Consistency with the direct measurements}

   While the limited $R_{eff}$ ($\lesssim 2$~kpc) seem to contradict the
claimed extended sizes of high-z SMGs (see \S 1), there is a reconciliation
when considering that even the sub-arcsec sub-mm interferometry might still
cannot resolve individual DSBRs at high-z ($0.1\arcsec \sim 0.8$~kpc at
$z\approx 1-4$) if they are too close to each other. In other words, the very
extended sub-mm morphologies could simply be due to the blending of a number of
discrete DSBRs that are widely separated but still not resolved by the
interferometry available today.

   Currently, the only way to resolve the individual DSBRs at high-z
is through strong gravitational lensing, in which case the highly
amplified background source could show its morphology in great details, albeit
in the distorted image plane. By properly modeling the lens, its image in the
source plane could be reconstructed to reveal the intrinsic morphology.
However, the accuracy of the reconstruction depends highly on how strong the
amplification is and how well the lens is modeled. 

   Unfortunately, there are still not many such measurements publicly available
for us to compare to, especially when we require spectroscopic
redshifts and the SPIRE photometry. The best example to date is the SMG
SMMJ2135-0102 from S\&I10, which is amplified by a factor
of $\sim 32.5$. Its FIR emission is dominated by four individual DSBRs, each
being mirrored into two images. By averaging the results from the
reconstructions based on the two sets of mirrored images, the intrinsic
radii of these four DSBRs are 390.0, 290.5, 192.5, and 94~pc, respectively.
As these radii are the FWHM sizes, and hence only enclose $\sim 76.1$\% of the
total light from each DSBR, they should be corrected by multiplying a factor
of $\sqrt{1/0.761}=1.146$ to the ``full sizes''. Following \S 3.2,
we get
$\sqrt{\sum\limits_{i} R_i^2}=609$~pc. In \S 3.1, we obtained
$L_{IR}^{(cd)}=\num{1.74d12}L_\odot$, $T_{mbb}=39.0$~K and
$R_{eff}=677$~pc. This size agrees with the above to $\sim 11$\%. Considering
that these four DSBRs, while being the dominant sources, might not contribute
100\% of the total $L_{IR}^{(cd)}$, the actual agreement could be even better.

   For comparison, here we also discuss the case of Arp 220, which is the
representative of the classic (low-redshift) ULIRG population. Arp 220 has two
nuclei, which have recently been directly resolved in sub-mm by
\citet{Scoville2015} using the Atacama Large Millimeter/submillimeter Array
(ALMA). Based on their highest resolution observations at \SI{870}\micron{}
(half-power bandwidth of $0.52\arcsec \times 0.39\arcsec$), these authors
derived the FWHM sizes of the two nuclei (assuming Gaussian profiles) as
$130\times 87$ and $137\times 116$~pc, respectively. Using spheres for
approximation, we obtain the equivalent radii of 106 and 126~pc, respectively.
They also suggest that these two nuclei account for $\sim 71$-76\% of the total
FIR continuum, and hence we adopt the middle value of 74\%. Similar to the
calculation above, we correct the measured FWHM sizes by a factor of
$\sqrt{1/(0.761\times 0.74)}=1.333$ to the full sizes, and obtain
$\sqrt{\sum\limits_{i} R_i^2}=219$~pc. As Arp 220 is very close
(we adopt 77~Mpc as its luminosity distance), using the SPIRE bands alone would
not result in a reasonable fit to its SED because these bands are too far away
from the peak of its FIR emission. Therefore, we use the photometry at
$>40$\micron, retrieved from the NASA/IPAC Extragalactic Database.
We obtain $L_{IR}^{(cd)}=\num{1.32d12}L_\odot$, $T_{mbb}=55.8$~K and
$R_{eff}=260$~pc. Considering the uncertainties (especially the fractional
contribution from the two nuclei to the total FIR emission), the agreement is
very reasonable.

\section{Discussion}

   Our MBB model fixes $\lambda_0$ to \SI{100}\micron, which is about the
smallest choice adopted in the literature.  As discussed in MY15 (see Appendix
A), the difference in $\lambda_0$ impacts the derived $T_{mbb}$
significantly but has little effect on $L_{IR}^{(cd)}$. If the SED fitting
can achieve the similar quality, adopting a larger
$\lambda_0$ will result in a higher $T_{mbb}$, and hence a smaller
$R_{eff}$. For a given $\lambda_0$, $R_{eff}$ still sensitively depends
on $T_{mbb}$. For example, if we take a crude approximation and ignore the
temperature dependence of $\tilde\sigma$, we have
$\Delta R_{eff}/R_{eff}\approx 2\times \Delta T_{mbb}/T_{mbb}$. 
Therefore, it is difficult to derive $R_{eff}$ to an accuracy
better than a factor of a few, especially at 
$L_{IR}^{(cd)}\lesssim 10^{11}L_\odot$ where the MBB S-B equivalent curves
are highly degenerated. However, our main conclusions still hold
regardless of such limitations.

   First of all, $R_{eff}$ cannot be increased arbitrarily. Even in the HyLIRG
regime, it is still mostly confined to $\lesssim 2$~kpc. While in the low
luminosity regime the increasing of $L_{IR}^{(cd)}$ can be achieved by
increasing $R_{eff}$ (e.g., increasing the number of DSBRs), in the high
luminosity regime this can only be achieved by increasing the strength
of the starburst (i.e., reflected in the rapid increase of $T_{mbb}$). MY15
has already reached this conclusion, and here we reinforce it with an
enlarged sample that consists of objects from different initial selections,
such as SMGs, high-z radio galaxies, etc. 

   Second, the $L_{IR}$--$T$ relation reflects the equivalent of the
Stefan-Boltzmann law in case of MBB and the aforementioned limit to $R_{eff}$.
For this reason, this relation outlines the boundary of minSFR3D,
which provides new clues in understanding dusty starbursting environment. The
value of minSFR3D should be quite close to the actual SFR volume density,
although strictly speaking it is only the lower limit. In the simplest case
where the DSBRs are all spheres of the same radius and dust temperature,
$\rho_\text{SFR}=\sqrt{N}\rho_\text{SFR}^{min}$, where $N$ is the total
number of DSBRs in the galaxy.

   Third, the small $R_{eff}$ values suggest that the DSBRs in high-z ULIRGs
are as physically compact as their counterparts in the local universe. It is
true that a small $R_{eff}$ could still be the result of an extended DSBR with
a low filling factor. However, our test cases in \S 3.4 strongly supports
the scenario that DSBRs are universally compact.

\acknowledgments{
  We acknowledge the support of the University of Missouri Research Board Grant
RB 15-22 and NASA's Astrophysics Data Analysis Program under grant number
NNX15AM92G. We have made use of the NASA/IPAC
Extragalactic Database (NED), which is operated by the Jet Propulsion
Laboratory, California Institute of Technology, under contract with
NASA\@.
}

\end{document}